\documentclass[
twocolumn,
unsortedaddress,
superscriptaddress,
showpacs,
pra
]{revtex4}

\usepackage{amsmath}
\usepackage{amssymb}
\usepackage{bm}
\usepackage{amsthm}

\newtheorem{theorem}{Theorem}
\newtheorem{lemma}{Lemma}

\newtheorem{proposition}{Proposition}

\newcommand{\Def}{:=}

\newcommand{\THEN}{\ensuremath{\,\Rightarrow\,}}

\newcommand{\IFF}{\ensuremath{\,\Leftrightarrow\,}} 
\newcommand{\wrt}{{w.r.t.~}}
\newcommand{\resp}{{resp.~}}

\newcommand{\C}{\ensuremath{{\mathcal{C}}}}

\newcommand{\E}{\ensuremath{{\mathcal{E}}}}
\newcommand{\F}{\ensuremath{{\mathcal{F}}}}

\renewcommand{\H}{\ensuremath{{\mathcal{H}}}}
\newcommand{\I}{\ensuremath{{\mathcal{I}}}}

\newcommand{\K}{\ensuremath{{\mathcal{K}}}}
\renewcommand{\L}{\ensuremath{{\mathcal{L}}}}
\newcommand{\M}{\ensuremath{{\mathcal{M}}}}

\newcommand{\R}{\ensuremath{{\mathcal{R}}}}
\renewcommand{\S}{\ensuremath{{\mathcal{S}}}}

\newcommand{\Comp}{\ensuremath{{\mathbb{C}}}} 

\renewcommand{\epsilon}{\varepsilon}
\renewcommand{\phi}{\varphi}

\newcommand{\etal}{\textit{et al.}~}

\DeclareMathOperator{\Tr}{Tr}

\newcommand{\oneover}[1]{\frac{1}{#1}}
\newcommand{\defset}[2]{\left\{#1\left|\,#2\right.\right\}}

\renewcommand{\Im}{\mathrm{Im}\,} 
\newcommand{\Ker}{\mathrm{Ker}\,} 

\newcommand{\cket}[1]{\left| #1 \right\rangle}
\newcommand{\bra}[1]{\left\langle #1 \right|}
\newcommand{\cketbra}[2]{%
\left| #1 \right\rangle\!\left\langle #2 \right|}
\newcommand{\pure}[1]{%
\left| #1 \right\rangle\!\left\langle #1 \right|}

\newcommand{\QO}{\ensuremath{{\mathcal{QO}}}}

\begin{document}

\title{
Perfect Quantum Error-Correcting Condition Revisited
}

\author{Tomohiro Ogawa}
\email{ogawa@mist.i.u-tokyo.ac.jp}
\affiliation{
Graduate School of Information Science and Technology,
University of Tokyo,
7-3-1 Hongo, Bunkyo-ku, Tokyo, 113-8656 Japan.
}

\date{\today}

\begin{abstract}
A simple and unifying method
to show the perfect error-correcting condition
is provided based on the quantum mutual information.
The one-to-one parameterization of quantum operations
and the properties of the quantum relative entropy
are used effectively in this paper,
where the equivalence between the subspace transmission
and the entanglement transmission is clearly presented.
We also revisit a variant of the no-cloning and no-deleting theorem
based on an information-theoretical tradeoff
between two parties for the reversibility of quantum operations,
and demonstrate that the no-cloning and no-deleting theorem
leads to the perfect error-correcting condition on Kraus operators.
\end{abstract}



\maketitle

\section{Introduction}

In the past decade, much progress has been made
in the theory of the quantum error-correcting codes
\cite{Shor-pra52:QECC,Steane-prl77:QECC,Calderbank-Shor-CSS-code,Steane-CSS-code,Gottesman-pra54:QECC,Calderbank-etal-pral78}
along with information-theoretical developments
\cite{Schumacher-pra96,Schumacher-Nielsen,Bennett-et-al,Knill-Laflamme,Bennet-DiVincenzo-Smolin-prl78,Barnum-Nielsen-Schumacher-pra57,Barnum-Smolin-Terhal-pra58,Barnum-Knill-Nielsen-IT2000}.
In particular, the perfect quantum error-correcting condition
by Schumacher and Nielsen \cite{Schumacher-Nielsen}
is of much importance,
providing insights on the role of the coherent information
\cite{Schumacher-Nielsen,Lloyd-pra55}.
Another approach to the perfect error-correcting condition
was also given independently
by Knill and Laflamme \cite{Knill-Laflamme}
and
by Bennett \etal \cite{Bennett-et-al}
to establish the algebraic condition on Kraus operators
of quantum operations.
The above mentioned results are already widely known
and, for example, one may find them in the textbook by
Nielsen and Chuang \cite{Nielsen-Chuang}.

On the other hand, Cerf and Adami
\cite{Cerf-Adami-prl79,Adami-Cerf-pra56}
introduced a quantum counterpart of
the classical mutual information,
namely the \textit{quantum mutual information},
and tried to develop the quantum information theory through it
\cite{Cerf-Adami-prl79,Adami-Cerf-pra56,Cerf-Cleve-pra56,Cerf-pra57},
while
the results by Cerf and Cleve \cite{Cerf-Cleve-pra56,Cerf-pra57}
concerning the perfect error-correcting condition
were dependent on those of Schumacher and Nielsen
\cite{Schumacher-Nielsen}
and the role of the quantum mutual information remained relatively unclear.
Later the quantum mutual information appeared in
the formula for the entanglement-assisted capacity
\cite{BSST-EA-Capacity-prl83,BSST-EA-Capacity-IT2002,Holevo-EA-Capacity-JMP2002}
of quantum channels.
Note that the Holevo information \cite{Holevo73}
and the coherent information \cite{Schumacher-Nielsen,Lloyd-pra55}
are also regarded as quantum counterparts of the classical mutual information,
with definite meanings as the capacities of quantum channels
for transmitting classical information
\cite{Holevo-Capacity,Schumacher-Westmoreland}
and quantum information
\cite{Shor-Capacity,Devetak-Capacity}, respectively.
In this paper, however,
we use the term quantum mutual information
as one introduced by Cerf and Adami.

The aim of this paper is to provide
a simple and unifying method
to show the perfect error-correcting condition
based on the quantum mutual information.
Our approach does not depend on
the results of Schumacher and Nielsen \cite{Schumacher-Nielsen},
but the one-to-one parameterization
\cite{Fujiwara-Algoet,Fujiwara-private-communication,Fujiwara-QIC,Choi-CP}
of quantum operations
and the properties \cite{Petz1988:SufficientChannels,Petz2003:MonotonicityRevisited,Hayden-Jozsa-Petz-Winter}
of the quantum relative entropy
are used effectively.
The arguments in this paper
will refine the earlier works
\cite{Cerf-Cleve-pra56,Cerf-pra57},
and shed light on the role of the quantum mutual information.
As an application of our method,
a variant \cite{Cleve-Gottesman-Lo}
of the no-cloning theorem
\cite{Wootters-Zurek,Dieks,Yuen,Barnum-et-al,Koashi-Imoto:No-cloning,Lindblad:No-cloning}
and the no-deleting theorem
\cite{Pati-Braunstein:Nature,Pati-Braunstein:quant-ph/0007121}
is recaptured
based on an information-theoretical tradeoff \cite{Cerf-pra57}
between two parties for the reversibility of quantum operations.
It is also demonstrated that
our approach immediately yields
the perfect error-correcting condition on Kraus operators
\cite{Knill-Laflamme,Bennett-et-al}.
The results themselves in this paper are not always new
and should be regarded as a refinement or a recast of the earlier works
\cite{Schumacher-Nielsen,Bennett-et-al,Knill-Laflamme,Cerf-Cleve-pra56,Cerf-pra57,Cleve-Gottesman-Lo,Hayden-Jozsa-Petz-Winter}.
The methods used here, however,
give us clear insights on the role of the quantum mutual information
related to the reversibility of the quantum operations.

\section{Definitions and the No-Cloning and No-Deleting Theorem}

Let $\H_A$, $\H_B$, and $\H_C$ be finite dimensional Hilbert spaces,
and let $\L(\H)$ denote the totality of linear operators
on a Hilbert space $\H$.
The totality of density operators is denoted by
\begin{align}
\S(\H)\Def\defset{\rho\in\L(\H)}{\rho^*=\rho\ge 0,\,\Tr[\rho]=1}.
\end{align}
The notion of the \textit{reversibility} and the \textit{vanishing property}
is defined for quantum operations as follows,
related to the quantum error-correcting schemes.
A quantum operation $\E:\S(\H_A)\rightarrow\S(\H_B)$,
which is a trace preserving and completely positive linear superoperator,
is called
\textit{reversible} with respect to (w.r.t.)
a subset $\S\subseteq \S(\H_A)$
if there exists a quantum operation $\R:\S(\H_B)\rightarrow\S(\H_A)$
such that $\forall\rho\in\S,\,\R\E(\rho)=\rho$.
A quantum operation $\E:\S(\H_A)\rightarrow\S(\H_B)$
is called \textit{vanishing} \wrt $\S\subseteq \S(\H_A)$
if there exists a density operator $\rho_0\in\S(\H_B)$
such that $\forall\rho\in\S,\,\E(\rho)=\rho_0$.

In the quantum error-correcting schemes,
a subspace $\K_A\subseteq\H_A$ is chosen
as a codebook to be protected from a quantum operation $\E$
so that $\E$ is reversible \wrt the set
\begin{align}
\S_1(\K_A)\Def\defset{\pure{\psi}\in\S(\H_A)}{\cket{\psi}\in\K_A}
\end{align}
of pure states with their eigenvectors included by $\K_A$.
In this case, we may say that
$\E$ is reversible \wrt the subspace $\K_A$ for simplicity.
It should be noted
that the reversibility and the vanishing property
of $\K_A$ are, respectively, equivalent to
those of the convex hull of $\S_1(\K_A)$, say $\S(\K_A)$,
which is the set of density operators
with their supports included by $\K_A$.

Using the subspace $\K_A$ as the codebook,
an arbitrary quantum state on a Hilbert space $\H_X$
with $\dim\H_X=\dim\K_A$
is transmitted over the quantum channel $\E$.
The encoding operation for this purpose is given
by the isometry encoding
\begin{align}
\C:\rho_X\in\S(\H_X)\mapsto \rho_A=V\rho_X V^*\in\S(\H_A),
\label{isometry-one-to-one}
\end{align}
where $V:\H_X\rightarrow\H_A$ is an isometry satisfying $\Im V=\K_A$.
Note that $\S(\K_A)$ defined above is written
specifically using the isometry encoding as
\begin{align*}
\S(\K_A)=\defset{\rho_A\in\S(\H_A)}{\rho_A=V\rho_X V^*,\,\rho_X\in\S(\H_X)}.
\end{align*}
The point here is that
we have the one-to-one correspondence
\eqref{isometry-one-to-one} between $\S(\K_A)$ and $\S(\H_X)$
and
that any $\rho_A\in\S(\K_A)$ may be identified with some $\rho_X\in\S(\H_X)$.

A further definition is needed to state the following theorem.
A quantum operation $\E:\S(\H_A)\rightarrow\S(\H_B)$
is called a \textit{pure state channel} \wrt a subspace $\K_A\subseteq\H_A$
if the output $\E(\pure{\psi})$ of any pure state $\pure{\psi}$
in $\S_1(\K_A)$ results in a pure state.
The following theorem represents the tradeoff between two parties
for the reversibility of quantum operations
and is regarded as a variant of the no-cloning theorem
\cite{Wootters-Zurek,Dieks,Yuen,Barnum-et-al,Koashi-Imoto:No-cloning,Lindblad:No-cloning}
and the no-deleting theorem
\cite{Pati-Braunstein:Nature,Pati-Braunstein:quant-ph/0007121}.
Originally the following theorem was
shown by Cleve \etal \cite{Cleve-Gottesman-Lo}
in the case $\K_A=\H_A$
with explicit arguments for the proof
using the perfect error-correcting condition on Kraus operators
\cite{Bennett-et-al,Knill-Laflamme}.
One may find a rigorous description \cite{Ogawa-QSSS} of the proof
according to the original arguments \cite{Cleve-Gottesman-Lo}.
\begin{theorem}
\label{Theorem:no-cloning-deleting}
Given a quantum operation
$\E_{BC}:\S(\H_A)\rightarrow\S(\H_B\otimes\H_C)$,
let $\E_B\Def\Tr_C\E_{BC}$ and $\E_C\Def\Tr_B\E_{BC}$
be composite maps of quantum operations.
Then, concerning the following conditions
for a subspace $\K_A\subseteq\H_A$,
\begin{enumerate}
\item[(i)]
$\E_B$ is reversible \wrt $\S(\K_A)$,
\item[(ii)] $\E_C$ is vanishing \wrt $\S(\K_A)$,
\end{enumerate}
it holds that
\begin{enumerate}
\item[(a)]
(i) \THEN (ii),
\item[(b)]
(i) \!\IFF (ii)
if $\E_{BC}$ is a pure state channel \wrt $\K_A$
and reversible \wrt $\S(\K_A)$.
\end{enumerate}
\end{theorem}
In the later sections,
we provide a simple proof of the above theorem
using an information-theoretical tradeoff \cite{Cerf-pra57}
of the quantum mutual information between two parties.
The process to show the proof will expose the role
of the quantum mutual information
related to the reversibility of quantum operations.

\section{One-to-One Parameterization of Quantum Operations}

The aim of this section is, for readers' convenience,
to summarize the one-to-one parameterization
of quantum operations given by Fujiwara and Algoet
\cite{Fujiwara-Algoet,Fujiwara-private-communication}
(see also \cite{Fujiwara-QIC})
based on the work of Choi \cite{Choi-CP}.
The parameterization establishes an one-to-one affine correspondence
between the totality of quantum operations,
\begin{align*}
\QO
&\Def\defset{\E:\S(\H_A)\rightarrow\S(\H_B)}{\E:\text{quantum operation}},
\end{align*}
and a set of nonnegative definite operators which is defined below.

Let $d\Def\dim\H_A$ and
\begin{align}
\cket{\Phi}\Def\oneover{\sqrt{d}}\sum_{i=1}^d \cket{i}\otimes\cket{i}
\end{align}
be the standard maximally entangled state
on a bipartite system $\H_R\otimes\H_A$
with a reference system $\H_R$ satisfying $\dim\H_R=\dim\H_A$,
where $\{\cket{i}\}_{i=1}^d$ is a complete orthonormal basis
on $\H_A$, and we use the same index for that on $\H_R$ for simplicity.
Let us consider the output of the maximally entangled state
by the extended quantum operation $\I_R\otimes\E$
with $\I_R$ denoting the identity superoperator, i.e.,
\begin{align}
M(\E)&\Def(\I_R\otimes\E)(\cketbra{\Phi}{\Phi}).
\label{QO-parametrization}
\end{align}
On the other hand, define the set of nonnegative definite
operators on the extended Hilbert space by
\begin{align*}
\M\Def\defset{M\in\L(\H_R\otimes\H_B)}{M\ge 0,\Tr_B[M]=\oneover{d}I_R}.
\end{align*}
Then, the map
$\E\in\QO\mapsto M(\E)\in\M$
establishes the one-to-one affine correspondence
between $\QO$ and $\M$.

The essence of the one-to-one parameterization
can be seen from the Kronecker product representation of $M(\E)$,
\begin{align}
M(\E)
&=\oneover{d}\sum_{i=1}^d\sum_{j=1}^d\cketbra{i}{j}\otimes\E(\cketbra{i}{j})
\nonumber \\
&\simeq
\oneover{d}
\begin{pmatrix}
&& \\
&\E(\cketbra{i}{j})& \\
&&
\end{pmatrix}_{ij},
\label{ChoiMatrix}
\end{align}
which means the block matrix
including $\E(\cketbra{i}{j})$ in $(i,j)$-block.
Here note that the quantum operation defined
in $\S(\H_A)$ is naturally extended to $\L(\H_A)$
by the linearity and the polar identity,
\begin{align}
\cketbra{i}{j}
=\oneover{2}\left\{
\pure{a}-\pure{b}+\sqrt{-1}(\pure{c}-\pure{d})
\right\},
\label{polar-identity}
\end{align}
where
$\cket{a}=(\cket{i}+\cket{j})/\sqrt{2}$,
$\cket{b}=(\cket{i}-\cket{j})/\sqrt{2}$,
$\cket{c}=(\cket{i}+\sqrt{-1}\cket{j})/\sqrt{2}$,
and $\cket{d}=(\cket{i}-\sqrt{-1}\cket{j})/\sqrt{2}$.
Then we can see that
\eqref{ChoiMatrix} has the entire information about $\E$,
since the output $\E(\cketbra{i}{j})$ of each
of the complete basis 
$\{\cketbra{i}{j}\}_{ij}$ on the linear space $\L(\H_A)$
appears in the $(i,j)$-block of $M(\E)$;
see Appendix for details.

In the same way,
we can also make another one-to-one affine parameterization
from an arbitrary faithful state
\cite{Fujiwara-private-communication}.
Let $\rho_A>0$ be a faithful state in $\S(\H_A)$ and
\begin{align}
\rho_A= \sum_{i=1}^d p_i\cketbra{i}{i}
\end{align}
be the Schatten decomposition of $\rho_A$,
where $p_i$ is the eigenvalue
corresponding to the eigenvector $\cket{i}$.
Then, a purification of $\rho_A$ is given by
\begin{align}
\cket{\Phi_{\rho_A}}\Def\sum_{i=1}^d\sqrt{p_i}\cket{i}\otimes\cket{i}
\in\H_R\otimes\H_A.
\end{align}
Let $\rho_{RA}\Def\pure{\Phi_{\rho_A}}$ and
$\rho_R\Def\Tr_A[\rho_{RA}]$,
and define the output of the state $\rho_{RA}$
by the extended quantum operation as
\begin{align}
M_{\rho_{RA}}(\E)\Def(\I_R\otimes\E)(\rho_{RA}).
\label{general-one-to-one}
\end{align}
On the other hand,
let us define a set of nonnegative definite operators by
\begin{align*}
\M_{\rho_{RA}}\Def\defset{M\in\L(\H_R\otimes\H_B)}{M\ge 0,\Tr_B[M]=\rho_R}.
\end{align*}
Then the map
$\E\in\QO \mapsto M_{\rho_{RA}}(\E)\in\M_{\rho_{RA}}$, again,
establishes the one-to-one affine correspondence between $\QO$
and $\M_{\rho_{RA}}$.
\begin{widetext}
This fact is also verified by the Kronecker product representation,
\begin{align}
M_{\rho_{RA}}(\E)
=
\sum_{i=1}^d\sum_{j=1}^d
\sqrt{p_i}\sqrt{p_j}\cketbra{i}{j}\otimes\E(\cketbra{i}{j})
\simeq
\begin{pmatrix}
\sqrt{p_1}I_B& & 0 \\
&\ddots& \\
0 && \sqrt{p_d}I_B
\end{pmatrix}
\begin{pmatrix}
&& \\
&\E(\cketbra{i}{j})& \\
&&
\end{pmatrix}_{ij}
\begin{pmatrix}
\sqrt{p_1}I_B& & 0 \\
&\ddots& \\
0 && \sqrt{p_d}I_B
\end{pmatrix},
\end{align}
\end{widetext}
which implies the one-to-one affine correspondence
between $M(\E)$ and $M_{\rho_{RA}}(\E)$.
Combining this with the correspondence between $\E$ and $M(\E)$
leads to the one-to-one affine correspondence
between $\E$ and $M_{\rho_{RA}}(\E)$.

\section{The Quantum Relative Entropy}

Let us define the quantum relative entropy
between two quantum states $\rho,\sigma\in\S(\H_A)$ by
\begin{align}
D(\rho||\sigma)\Def\Tr[\rho(\log\rho-\log\sigma)].
\end{align}
Then, for any quantum operation $\E:\S(\H_A)\rightarrow\S(\H_B)$,
it holds that
\begin{align}
D(\rho||\sigma)\ge D(\E(\rho)||\E(\sigma)),
\label{quantum-relative-entropy-monotonicity}
\end{align}
which is called
the monotonicity
\cite{Lindblad:CP,Uhlmann:CP}
and is one of the most important properties of the quantum relative entropy.
It is known that the equality of the monotonicity
\eqref{quantum-relative-entropy-monotonicity} holds
iff $\E$ is reversible \wrt $\{\rho,\sigma\}$
\cite{Petz1988:SufficientChannels}
(see also
Refs.~\cite{Petz2003:MonotonicityRevisited,Hayden-Jozsa-Petz-Winter}).
When the equality of the monotonicity holds, there is
a canonical reverse operation depending only on $\sigma$,
which given by
\begin{align}
\R_\sigma(\tau)
\Def\sigma^{\frac{1}{2}}
\E^*(\E(\sigma)^{-\frac{1}{2}}\tau\E(\sigma)^{-\frac{1}{2}})
\sigma^{\frac{1}{2}}
\label{canonical-reverse}
\end{align}
on the support of $\E(\sigma)$.
Here $\E^*:\L(\H_B)\rightarrow\L(\H_A)$ is the dual of $\E$ satisfying
\begin{align*}
\forall \rho\in\S(\H_A),\,\forall Y\in\L(\H_B),
\,\Tr[\E(\rho)Y]=\Tr[\rho\E^*(Y)].
\end{align*}
The above fact is summarized as the follows.
\begin{proposition}[Petz \cite{Petz1988:SufficientChannels,Petz2003:MonotonicityRevisited,Hayden-Jozsa-Petz-Winter}]
\label{Prop:divergence-equality}
Given a quantum operation $\E:\S(\H_A)\rightarrow S(\H_B)$
and $\rho,\sigma\in\S(\H_A)$,
let $\R_\sigma$ be the quantum operation defined by \eqref{canonical-reverse}.
Then the following conditions are equivalent.
\begin{enumerate}
\item[(a)] $D(\rho||\sigma) = D(\E(\rho)||\E(\sigma))$.
\item[(b)] $\R_{\sigma}\E(\rho)=\rho$.
\item[(c)] $\E$ is reversible \wrt $\{\rho,\sigma\}$.
\end{enumerate}
\end{proposition}

The quantum relative entropy also satisfies other important properties.
One of them is the positivity,
\begin{align}
D(\rho||\sigma)\ge 0,
\quad D(\rho||\sigma)=0 \IFF \rho=\sigma.
\label{divergence-positivity}
\end{align}
Another one is
the invariance under the action of unitary transformations or isometries,
i.e.,
it holds for any isometry $V:\H_A\rightarrow\H_B$ that
\begin{align}
D(\rho||\sigma)=D(V\rho V^*||V\sigma V^*).
\label{divergence-unitary-invariance}
\end{align}

\section{The Quantum Mutual Information and the Reversibility}

Let us define the quantum mutual information
\cite{Cerf-Adami-prl79,Adami-Cerf-pra56}
for a bipartite state $\rho_{XY}\in\S(\H_X\otimes\H_Y)$ by
\begin{align}
I_{\rho_{XY}}(X;Y)\Def H(X)+H(Y)-H(XY),
\end{align}
where $H(X)$, $H(Y)$, and $H(XY)$ are the von Neumann entropy,
$H(\rho)\Def-\Tr[\rho\log\rho]$,
of the corresponding states
$\rho_X=\Tr_Y\rho_{XY}$, $\rho_Y=\Tr_X\rho_{XY}$,
and $\rho_{XY}$, respectively.
Hereafter, the subscript $\rho_{XY}$ is omitted
if the state is fixed and no confusion is likely to arise.
It is widely known that
the quantum mutual information is also written as
\begin{align}
I_{\rho_{XY}}(X;Y)=D(\rho_{XY}||\rho_X\otimes\rho_Y).
\end{align}

We shall introduce an information-theoretical quantity $I(\rho_A,\E)$
\cite{Adami-Cerf-pra56},
defined with respect to
a quantum operation $\E:\S(\H_A)\rightarrow\S(\H_B)$
and a density operator $\rho_A\in\S(\H_A)$,
which plays an important role in this paper.
Let $\cket{\Psi_{\rho_A}}\in\H_R\otimes\H_A$ be
a purification of $\rho_A$ with a reference system $\H_R$, and
\begin{align}
\rho_{RA}&\Def\pure{\Psi_{\rho_A}},
\\
\rho_{RB}&\Def(\I_R\otimes\E)(\rho_{RA}).
\end{align}
Then $I(\rho_A,\E)$ is defined by
\begin{align}
I(\rho_A,\E)&\Def I_{\rho_{RB}}(R;B)
\nonumber \\
&=D((\I_R\otimes\E)(\rho_{RA})||(\I_R\otimes\E)(\rho_R\otimes\rho_A)).
\label{mutual-info-channel}
\end{align}
Note that the freedom in purifications
is essentially described by an isometry $V_R$ acting on $\H_R$
so that another purification is given by
$\rho_{R'A}=(V_R\otimes\I_A)\rho_{RA}(V_R\otimes\I_A)^*$.
Hence, we can see that \eqref{mutual-info-channel}
is independent of a specific realization of purifications,
since the quantum relative entropy is kept invariant
under the action of
isometries as described in \eqref{divergence-unitary-invariance}.

It is also clear that the monotonicity of the quantum relative entropy
\eqref{quantum-relative-entropy-monotonicity}
yields the data processing inequality
\cite{Adami-Cerf-pra56}
for the quantum mutual information \eqref{mutual-info-channel},
\begin{align}
I(\rho_A,\I_A)\ge I(\rho_A,\E)\ge I(\rho_A,\F\E),
\label{mutual-info-monotonicity}
\end{align}
where $\F:\S(\H_B)\rightarrow\S(\H_C)$
is an arbitrary further quantum operation.
From Proposition \ref{Prop:divergence-equality},
the first equality of \eqref{mutual-info-monotonicity}
holds iff
\begin{align}
\R_{\rho_R\otimes\rho_A}(\I_R\otimes\E)(\rho_{RA})=\rho_{RA},
\end{align}
where $\R_{\rho_R\otimes\rho_A}$ is the reverse operation
defined in \eqref{canonical-reverse}.
Note that the above condition is also written as
\begin{align}
(\I_R\otimes\R_{\rho_A})(\I_R\otimes\E)(\rho_{RA})=\rho_{RA},
\label{mutal-info-rev-cond}
\end{align}
which is demonstrated by Hayden \etal \cite{Hayden-Jozsa-Petz-Winter}.
They discussed the equivalence between
the first equality of \eqref{mutual-info-monotonicity}
and \eqref{mutal-info-rev-cond}
with an explicit construction of the recovery operation.
Now we have the following lemma, which is essential for later discussions.
\begin{lemma}
\label{Lemma:reversible-vanishing-condition}
For a quantum operation $\E:\S(\H_A)\rightarrow\S(\H_B)$,
the following three conditions
on the reversibility and the vanishing property
are, respectively, equivalent.
\begin{enumerate}
\item[(a)]
$\E$ is reversible (\resp vanishing) \wrt $\S(\H_A)$.
\item[(b)]
$\forall\rho_A\in\S(\H_A),\,I(\rho_A,\E)=I(\rho_A,\I_A)$ (\resp $=0$).
\item[(c)]
$\exists\rho_A>0,\,I(\rho_A,\E)=I(\rho_A,\I_A)$ (\resp $=0$).
\end{enumerate}
\end{lemma}
\begin{proof}
The equivalence of the above conditions for the reversibility
is shown as follows.

\noindent
(a)\THEN(b):
If $\E$ is reversible \wrt $\S(\H_A)$, then there exists
a quantum operation $\R:\S(\H_B)\rightarrow\S(\H_A)$
such that $\R\E=\I_A$.
Then letting $\F=\R$ in
\eqref{mutual-info-monotonicity}, we have (b).

\noindent
(b)\THEN(c): Obvious.

\noindent
(c)\THEN(a):
For the state $\rho_A>0$ given in (c), we have
\begin{align}
&I(\rho_A,\E)=I(\rho_A,\I_A)
\nonumber \\
\IFF
&(\I_R\otimes\R_{\rho_A})(\I_R\otimes\E)(\rho_{RA})=\rho_{RA}
\label{rev-cond-2} \\
\IFF
&\R_{\rho_A}\E=\I_A,
\label{rev-cond-3}
\end{align}
where \eqref{rev-cond-2} follows from
Proposition \ref{Prop:divergence-equality} and
\eqref{mutal-info-rev-cond},
and \eqref{rev-cond-3} follows from the one-to-one parameterization
\eqref{general-one-to-one}.
Now \eqref{rev-cond-3} implies (a).

Next, we turn to the vanishing condition.

\noindent
(a)\THEN(b):
If $\E$ is vanishing \wrt $\S(\H_A)$, then
it is nothing but the composition map of
the trace operation on $\H_A$ and
the creation of a state $\rho_0\in\S(\H_B)$.
Therefore, we have
\begin{align}
(\I_R\otimes\E)(\rho_{RA})=\rho_R\otimes\rho_0,
\end{align}
which implies (b).

\noindent
(b)\THEN(c): Obvious.

\noindent
(c)\THEN(a):
For the state $\rho_A>0$ given in (c),
it follows from \eqref{divergence-positivity} that
\begin{align}
&I(\rho_A,\E)=0
\IFF
(\I_R\otimes\E)(\rho_{RA})=\rho_R\otimes\E(\rho_A).
\label{rev-cond-4}
\end{align}
Thus, the one-to-one parameterization
\eqref{general-one-to-one} of $\E$
coincides with that of the composition map of
the trace operation and the creation of the state $\E(\rho_A)$,
which implies (a).
\end{proof}

Considering the isometry encoding
discussed in \eqref{isometry-one-to-one},
Lemma \ref{Lemma:reversible-vanishing-condition}
is strengthened as follows.
\begin{theorem}
\label{Theorem:error-correcting-condition}
For a quantum operation $\E:\S(\H_A)\rightarrow\S(\H_B)$
and a subspace $\K_A\subseteq\H_A$,
the following three conditions
on the reversibility and the vanishing property
are, respectively, equivalent.
\begin{enumerate}
\item[(a)]
$\E$ is reversible (\resp vanishing) \wrt $\S(\K_A)$.
\item[(b)]
$\forall\rho_A\in\S(\K_A),\,I(\rho_A,\E)=I(\rho_A,\I_A)$ (\resp $=0$).
\item[(c)]
There exists a density operator $\rho_A$ with its support $\K_A$ such that
$I(\rho_A,\E)=I(\rho_A,\I_A)$ (\resp $=0$).
\end{enumerate}
\end{theorem}
\begin{proof}
Consider the isometry encoding $\C$
discussed in \eqref{isometry-one-to-one}
to send quantum states on a Hilbert space $\H_X$
with its dimension $\dim\H_X=\dim\K_A$.
Then, any $\rho_A\in\S(\K_A)$ is identified with
a state $\rho_X\in\S(\H_X)$ by the one-to-one correspondence
$\rho_A=\C(\rho_X)$, and it holds that
\begin{align}
I(\rho_X,\I_X)&=I(\rho_X,\C)=I(\rho_A,\I_A),
\label{isometry-encoding-1}
\\
I(\rho_X,\E\C)&=I(\rho_A,\E),
\label{isometry-encoding-2}
\end{align}
where the first equality of \eqref{isometry-encoding-1}
follows from the reversibility of $\C$
and Lemma \ref{Lemma:reversible-vanishing-condition},
and the other equalities follow from the one-to-one correspondence
$\rho_A=\C(\rho_X)$.
Applying Lemma \ref{Lemma:reversible-vanishing-condition}
with $\E$ and $\H_A$ replaced with $\E\C$ and $\H_X$, respectively,
we obtain the assertion.
\end{proof}

As for the reversible conditions,
Theorem \ref{Theorem:error-correcting-condition}
is just a recast of the famous perfect error-correcting condition
by Schumacher and Nielsen \cite{Schumacher-Nielsen}
using the coherent information,
\begin{align}
I_c(\rho_A,\E)\Def H(B)-H(RB).
\label{coherent-info-def}
\end{align}
They proved that
\begin{align}
I_c(\rho_A,\E)=H(\rho_A)
\label{coherent-info-equal}
\end{align}
holds iff there exists a quantum operation
$\R:\S(\H_B)\rightarrow\S(\H_A)$
such that
\begin{align}
(\I_R\otimes\R\E)(\rho_{RA})=\rho_{RA}.
\label{coherent-info-rev-cond}
\end{align}
The equivalence of their condition \eqref{coherent-info-equal} and
Theorem \ref{Theorem:error-correcting-condition}
is verified by
\begin{align}
I(\rho_A,\E)&=H(R)+I_c(\rho_A,\E),
\label{coherent-info-1}
\\
I(\rho_A,\I_A)&=H(R)+H(\rho_A),
\label{coherent-info-2}
\end{align}
where \eqref{coherent-info-2} follows from $H(RA)=0$
(see Ref.~\cite{Hayden-Jozsa-Petz-Winter}).
It is interesting to observe
from Theorem \ref{Theorem:error-correcting-condition},
\eqref{coherent-info-1}, and $H(R)=H(A)$
that the vanishing conditions are equivalent to
\begin{align}
I_c(\rho_A,\E)=-H(\rho_A).
\end{align}
Note that
the quantum data processing inequality
\cite{Schumacher-Nielsen}
for the coherent information
follows from \eqref{mutual-info-monotonicity}, i.e.,
\begin{align}
H(\rho_A)\ge I_c(\rho_A,\E)\ge I_c(\rho_A,\F\E).
\label{coherent-info-monotonicity}
\end{align}

Using relations \cite{Schumacher-pra96,Knill-Laflamme} between
the entanglement fidelity and the average fidelity,
Schumacher and Nielsen \cite{Schumacher-Nielsen}
also showed the equivalence between the subspace transmission
and the entanglement transmission, that is,
the equivalence between the condition (a)
in Theorem \ref{Theorem:error-correcting-condition}
and the condition \eqref{coherent-info-rev-cond} for some $\rho_A\in\S(\K_A)$.
One of the findings in Theorem \ref{Theorem:error-correcting-condition}
lies in a simple exposition of
the equivalence between them,
as well as providing the clear approach to treat
the perfect error-correcting condition.
In our approach,
the meaning of the entanglement fidelity,
\begin{align}
F_e(\rho_A,\R\E)
\Def
\bra{\Phi_{\rho_A}}(\I_R\otimes\R\E)(\pure{\Phi_{\rho_A}})\cket{\Phi_{\rho_A}},
\label{entanglement-fidelity-def}
\end{align}
should be translated into the measure
of how close the quantum operation $\R\E$
is to the identity operation $\I_A$,
at least on the support of $\rho_A$,
in the sense of the one-to-one affine parameterization
\eqref{general-one-to-one}
\cite{Fujiwara-Algoet,Fujiwara-private-communication}.

It is remarked that, in Ref.~\cite{Anderson},
the condition (c) on the vanishing property
in Lemma \ref{Lemma:reversible-vanishing-condition}
was used as the unauthorized condition
for quantum secret sharing schemes
\cite{Hillery-et-al,Karlson-et-al,Cleve-Gottesman-Lo}.

\section{Tradeoff between Two Parties}

In this section, we explore the tradeoff between two parties
for the reversibility of quantum operations,
and provide
a simple proof of Theorem \ref{Theorem:no-cloning-deleting}.
The tradeoff for the reversibility
can be clearly seen through the following theorem.
Originally, the equality \eqref{info-preserv} in the following theorem
was given by Cerf \cite{Cerf-pra57}
in the case of error-correcting schemes.
\begin{theorem}
\label{Theorem:mutual-info-preserving}
Given a quantum $\E_{BC}:\S(\H_A)\rightarrow\S(\H_B\otimes\H_C)$,
let $\E_B\Def\Tr_C\E_{BC}$ and $\E_C\Def\Tr_B\E_{BC}$.
\begin{enumerate}
\item[(a)]
It holds for any $\rho_A\in\S(\H_A)$ that
\begin{align}
I(\rho_A,\I_A) \ge I(\rho_A,\E_{B})+I(\rho_A,\E_{C}).
\label{info-inequality}
\end{align}
\item[(b)]
If $\E_{BC}$ is a pure state channel \wrt a subspace $\K_A\subseteq\H_A$
and reversible \wrt $\S(\K_A)$,
then it holds for any $\rho_A\in\S(\K_A)$ that
\begin{align}
I(\rho_A,\I_A)
&=I(\rho_A,\E_{BC})
\nonumber \\
&=I(\rho_A,\E_{B})+I(\rho_A,\E_{C}).
\label{info-preserv}
\end{align}
\end{enumerate}
\end{theorem}
\begin{proof}
First, we show the assertion (b).
The first equality of \eqref{info-preserv}
follows from the reversibility of $\E_{BC}$ and
``(a)\THEN(b)'' of
Theorem \ref{Theorem:error-correcting-condition}.
If $\E_{BC}$ is a pure state channel \wrt $\K_A$
and reversible \wrt $\S(\K_A)$,
the output of $\rho_A\in\S(\K_A)$
is written as $\E_{BC}(\rho_A)=W\rho_A W^*$
by a partial isometry $W:\H_A\rightarrow\H_B\otimes\H_C$
satisfying $(\Ker W)^{\perp}=\K_A$.
Therefore the output of the purification,
\begin{align}
\rho_{RBC}
&\Def(\I_R\otimes\E_{BC})(\pure{\Phi_{\rho_A}})
\nonumber \\
&=(\I_R\otimes W)\pure{\Phi_{\rho_A}}(\I_R\otimes W)^*,
\label{info-preserv-1}
\end{align}
is also a pure state.
Then the purity of $\rho_{RBC}$ yields
the following equality \cite{Cerf-pra57},
\begin{align}
2H(R)&=I(R;B)+I(R;C),
\label{info-preserv-2}
\end{align}
which is verified
by using $H(B)=H(RC)$ and $H(RB)=H(C)$ as
\begin{align}
&I(R;B)+I(R;C)
\nonumber \\
&=H(R)+H(B)-H(RB)+H(R)+H(C)-H(RC)
\nonumber \\
&=2H(R).
\end{align}
On the other hand, we have
\begin{align}
I(R;BC)
&=H(R)+H(BC)-H(RBC)
\nonumber \\
&=2H(R),
\label{info-preserv-3}
\end{align}
which follows from
$H(R)=H(BC)$ and $H(RBC)=0$.
Now \eqref{info-preserv} follows from
\eqref{info-preserv-2} and \eqref{info-preserv-3}.

The assertion (a) is shown as follows.
let us consider the Stinespring dilation \cite{Stinespring}
$\E_{BC}(\rho_A)=\Tr_E[V\rho_AV^*]$,
where $V$ is an isometry from $\H_A$
to the composite system of $\H_B$$, \H_C$,
and an environment system $\H_E$.
Let $\E_{BCE}(\rho_A)\Def V\rho_AV^*$,
then the quatum operation $\E_{BCE}$ is
a pure state channel \wrt $\H_A$ and reversible \wrt $\S(\H_A)$.
Therefore, applying the above arguments, we have
\begin{align}
I(\rho_A,\I_A)
&= I(\rho_A,\E_{B})+I(\rho,\E_{CE})
\nonumber \\
&\ge I(\rho_A,\E_{B})+I(\rho,\E_{C}),
\end{align}
where the last inequality follows from the monotonicity
\eqref{mutual-info-monotonicity}.
\end{proof}

A proof of Theorem \ref{Theorem:no-cloning-deleting}
immediately follows from Theorem \ref{Theorem:error-correcting-condition}
and Theorem \ref{Theorem:mutual-info-preserving} as follows.
First, we show the assertion (b) of Theorem \ref{Theorem:no-cloning-deleting}.
Let us take an arbitrary state $\rho_A\in\S(\K_A)$
which has the support $\K_A$.
Suppose that $\E_{BC}$ is
a pure state channel \wrt $\K_A$ and reversible \wrt $\S(\K_A)$.
Then we have
\begin{align}
&\text{$\E_B$ is reversible \wrt $\S(\K_A)$}
\nonumber \\
\IFF
&I(\rho_A,\E_B)=I(\rho_A,\I_A)
\label{no-cloning-proof-1}
\\
\IFF
&I(\rho_A,\E_C)=0
\label{no-cloning-proof-2}
\\
\IFF
&\text{$\E_C$ is vanishing \wrt $\S(\K_A)$},
\label{no-cloning-proof-3}
\end{align}
where
\eqref{no-cloning-proof-1} and \eqref{no-cloning-proof-3} follow from
Theorem \ref{Theorem:error-correcting-condition},
and \eqref{no-cloning-proof-2} follows from \eqref{info-preserv}.

The assertion (a) of Theorem \ref{Theorem:no-cloning-deleting}
is shown in the same way as
\begin{align}
&\text{$\E_B$ is reversible \wrt $\S(\K_A)$}
\nonumber \\
\IFF
&I(\rho_A,\E_B)=I(\rho_A,\I_A)
\label{no-cloning-proof-4}
\\
\THEN
&I(\rho_A,\E_C)\le 0
\label{no-cloning-proof-5}
\\
\IFF
&I(\rho_A,\E_C)=0
\label{no-cloning-proof-6}
\\
\IFF
&\text{$\E_C$ is vanishing \wrt $\S(\K_A)$},
\label{no-cloning-proof-7}
\end{align}
where
\eqref{no-cloning-proof-5} follows from \eqref{info-inequality},
and \eqref{no-cloning-proof-6} follows from
the positivity \eqref{divergence-positivity}
of the quantum relative entropy.
It was also shown in Ref.~\cite{Anderson} that
\eqref{no-cloning-proof-4} implies \eqref{no-cloning-proof-6}
by a different method.

\section{Error-Correcting Condition on Kraus Operators}

We shall demonstrate that
Theorem \ref{Theorem:no-cloning-deleting} immediately yields
the quantum error-correcting condition on Kraus operators
which is proved independently
by Knill and Laflamme \cite{Knill-Laflamme}
and
by Bennett \etal \cite{Bennett-et-al}.
\begin{proposition}[\cite{Knill-Laflamme,Bennett-et-al}]
\label{Knill-Laflamme-Theorem}
Let $\E:\S(\H_A)\rightarrow\S(\H_B)$ be a quantum operation
represented by
the Kraus representation \cite{Kraus},
\begin{align}
\E(\rho)=\sum_kE_k\rho E_k^*,
\end{align}
and $\K_A\subseteq\H_A$ be a subspace.
Then the following conditions are equivalent,
where
the projection onto the subspace $\K_A$ is denoted by $P_{\K_A}$.
\begin{enumerate}
\item[(a)] $\E$ is reversible \wrt $\S(\K_A)$.
\item[(b)] For each pair of indices $(k,l)$,
there exists $c_{kl}\in\Comp$ such that
$
P_{\K_A} E_k^*E_l P_{\K_A}=c_{kl}P_{\K_A}
$.
\end{enumerate}
\end{proposition}
\begin{proof}
Let us consider the Stinespring dilation
$\E(\rho_A)=\Tr_E[V\rho_A V^*]$,
where $V$ is an isometry from $\H_A$
to the composite system of $\H_B$ and an environment system $\H_E$,
i.e.,
\begin{align}
V: \cket{i}\in\H_A \mapsto \sum_l E_l\cket{i}\otimes\cket{l}\in\H_B\otimes\H_E.
\label{Knill-Laflamme-10}
\end{align}
Let $\E_{BE}(\rho_A)\Def V\rho_A V^*$ and $\E_E\Def\Tr_B\E_{BE}$,
then the quantum operation $\E_{BE}$
is a pure state channel and reversible \wrt $\S(\H_A)$,
and hence,
it follows from Theorem \ref{Theorem:no-cloning-deleting}
that the condition (a) above
is equivalent to the condition,
\begin{enumerate}
\item[(c)] $\E_E$ is vanishing \wrt $\S(\K_A)$.
\end{enumerate}
We shall show that the condition (c)
is equivalent to the condition (b) above.
Using \eqref{Knill-Laflamme-10}, $\E_E$ is explicitly written as
\begin{align}
\E_E(\rho_A)=\sum_k\sum_l\Tr[\rho_AE_k^*E_l]\,\cketbra{l}{k}.
\label{Knill-Laflamme-30}
\end{align}
Therefore, (c) is equivalent to
\begin{align}
\forall(k,l),\,\exists c_{kl}\in\Comp,\forall\rho_A\in\S(\K_A),\,
\Tr[\rho_AE_k^*E_l]=c_{kl}.
\label{Knill-Laflamme-40}
\end{align}
Now let $\{\cket{i}\}_{i=1}^{\dim\K_A}$ be
a complete orthonormal basis on $\K_A$.
Then, by the polar identity \eqref{polar-identity},
we can show that
\eqref{Knill-Laflamme-40} is also equivalent to
the existence of $c_{kl}\in\Comp$ such that
\begin{align}
\bra{i}E_k^*E_l\cket{j} =c_{kl}\delta_{ij}
\quad (i,j=1,\dots,\dim\K_A),
\label{Knill-Laflamme-50}
\end{align}
which implies that the matrix components of
$P_{\K_A}E_k^*E_lP_{\K_A}$ are the same as those of $c_{kl}P_{\K_A}$.
Hence \eqref{Knill-Laflamme-50} is
equivalent to the condition (b) above.
\end{proof}

\section{Concluding Remarks}

We have laid a simple and unifying method
to show the perfect error-correcting condition
based on the quantum mutual information
(Theorem \ref{Theorem:error-correcting-condition}).
In our approach,
the one-to-one affine parameterization
of quantum operations
played an important role
to show the equivalence between the subspace transmission
and the entanglement transmission,
as well as the meaning of the entanglement fidelity
as the measure of closeness between a quantum operation
and the identity operation.
We have also revisited the no-cloning and no-deleting theorem
(Theorem \ref{Theorem:no-cloning-deleting})
based on the information-theoretical tradeoff
(Theorem \ref{Theorem:mutual-info-preserving})
between two parties for the reversibility of quantum operations,
and demonstrated that the no-cloning and no-deleting theorem
leads to the perfect error-correcting condition on Kraus operators.

In this paper, the study of the error-correcting schemes
was restricted to those with perfect reconstruction of
the encoded states,
while it is important to extend our approach
to the error-correcting schemes
allowing small errors or asymptotically vanishing errors.
In the process of extending our results,
the following pair of equalities given in Ref.~\cite{Devetak-Harrow-Winter},
\begin{align}
H(\rho_A)&=\oneover{2}\{I(\rho_A,\E)+I(\rho_A,\E_E)\},
\label{AE-tradeoff} \\
I_c(\rho_A,\E)&=\oneover{2}\{I(\rho_A,\E)-I(\rho_A,\E_E)\},
\label{AE-coherent-info}
\end{align}
will play a crucial role,
where \eqref{AE-tradeoff} is just the equality \eqref{info-preserv-2}
by Cerf \cite{Cerf-pra57} applied to the case in the proof of
Proposition \ref{Knill-Laflamme-Theorem},
and \eqref{AE-coherent-info} is verified in the same way.
Note that \eqref{AE-tradeoff} is closely related to
the reversibility of the quantum operation $\E$.
On the other hand, \eqref{AE-coherent-info} will
establish some relations between the quantum mutual information
and the coherent information related to the various capacities
\cite{BSST-EA-Capacity-prl83,BSST-EA-Capacity-IT2002,Holevo-EA-Capacity-JMP2002,Shor-Capacity,Devetak-Capacity}
of quantum operation $\E$.
These developments are given in the subsequent paper by the author.

\section*{Acknowledgment}

The author wishes to thank A.~Fujiwara
for teaching him the essence of the one-to-one affine parameterization
of quantum operations.
He is grateful to H.~Nagaoka and K.~Matsumoto
for their useful comments.

This research was partially supported
by the Ministry of Education, Science, Sports, and Culture,
Grant-in-Aid for Young Scientists (B), 17740050, 2005.

\appendix

\section*{Appendix}

The use of the one-to-one parameterization
\cite{Fujiwara-Algoet,Fujiwara-private-communication,Choi-CP}
(see also \cite{Fujiwara-QIC})
of quantum operations is essential in this paper.
In this appendix, for readers' convenience, we show that
the map $\E\in\QO\mapsto M(\E)\in\M$ defined in \eqref{QO-parametrization}
actually establishes the one-to-one affine parameterization.

First, the fact that $M(\E)\in\M$ if $\E\in\QO$
is verified as follows.
The first requirement $M(\E)\ge 0$ for $\M$ follows from
the complete positivity of the quantum operation $\E$,
and the second requirement $\Tr_B[M(\E)]=\oneover{d}I_R$
is a consequence of the trace preserving condition on $\E$.
Actually, we have
\begin{align}
\Tr_{B}[M(\E)]
\simeq
\oneover{d}
\begin{pmatrix}
&& \\
& \Tr[\E(\cketbra{i}{j})] & \\
&&
\end{pmatrix}_{ij}
\simeq
\oneover{d} I_R.
\end{align}
Injectivity of the map $M(\E)$ is clear
from the representation \eqref{ChoiMatrix}.
For $\E,\F\in\QO$ and $t\in[0,1]$, we have
\begin{align}
&M(t\E+(1-t)\F)
\nonumber \\
&=(I_R\otimes(t\E+(1-t)\F))(\pure{\Phi})
\nonumber \\
&=t(I_R\otimes\E)(\pure{\Phi})+(1-t)(I_R\otimes\F)(\pure{\Phi})
\nonumber \\
&=tM(\E)+(1-t)M(\F),
\end{align}
which shows that $M(\E)$ is an affine mapping.

Conversely, by taking the $(i,j)$-block of $M\in\M$ as
\begin{align}
\E_M(\cketbra{i}{j})=d\cdot\Tr_R[(\cketbra{i}{j}\otimes\I_B)^*M],
\end{align}
we obtain the corresponding quantum operation $\E_M$ from $M\in\M$.
In fact, we can show that $\E_M$
actually yields a quantum operation as follows.
From the condition $M\ge 0$ of $\M$,
we obtain the following decomposition
\begin{align}
M=\sum_k\pure{\Psi_k}
\label{Choi-matrix-10}
\end{align}
by using $\cket{\Psi_k}\in\H_R\otimes\H_B$,
for example,
which is given by the spectral decomposition of $M$.
Here let $\L(\H_A\rightarrow\H_B)$ be the totality of linear operators
from $\H_A$ to $\H_B$. Then the map
\begin{align}
E\in\L(\H_A\rightarrow\H_B)
\mapsto (I\otimes E)\cket{\Phi}\in\L(\H_R\otimes\H_B)
\label{Choi-matrix-20}
\end{align}
defines an one-to-one linear correspondence between
$\L(\H_A\rightarrow\H_B)$ and $\L(\H_R\otimes\H_B)$.
Therefore,
there exists an operator $E_k\in\L(\H_A\rightarrow\H_B)$ for each $k$
such that $\cket{\Psi_k}=(I_R\otimes E_k)\cket{\Phi}$,
and hence, \eqref{Choi-matrix-10} is written as follows
\begin{align}
M
&=\sum_k(I_R\otimes E_k)\pure{\Phi}(I_R\otimes E_k)^*
\nonumber \\
&=\oneover{d}\sum_{i=1}^d\sum_{j=1}^d
\cketbra{i}{j}\otimes\left(\sum_kE_k\cketbra{i}{j}E_k^*\right).
\end{align}
The above formula implies that $\E_M$ is represented
by the Kraus representation,
\begin{align}
\E_M(\rho_A)=\sum_kE_k\rho_AE_k^*,
\end{align}
which ensures the complete positivity of $\E_M$.
The trace preserving requirement for $\E_M$ easily follows
from $\Tr_B[M]=\oneover{d}I_R$, i.e.,
\begin{align}
\Tr\E_M(\cketbra{i}{j})
&=d\cdot\Tr_B\Tr_R[(\cketbra{i}{j}\otimes\I_B)^*M],
\nonumber \\
&=d\cdot\Tr_R(\cketbra{i}{j}\Tr_B[M])
\nonumber \\
&=\delta_{ij}.
\end{align}


\end{document}